\documentclass[aps,pra,showpacs, preprint]{revtex4}

\begin{document}

\title{Infinitely-long-range nonlocal potentials and the Bose-Einstein supersolid phase}

\author{Moorad Alexanian}
\email[]{alexanian@uncw.edu}

\affiliation{Department of Physics and Physical Oceanography\\
University of North Carolina Wilmington\\ Wilmington, NC
28403-5606\\}

\date{\today}

\begin{abstract}
It is shown, with the aid of the Bogoliubov inequality, that a Bose-Einstein condensate has the Bloch form and represents a self-organized supersolid provided the interaction between the condensate atoms is nonlocal and of infinitely long-range.
\end{abstract}


\maketitle {}

\section{Introduction} The existence of a Bose-Einstein condensate (BEC) in an ideal 3D quantum Bose gas served as a useful physical concept in the study of theoretical models for superfluidity, which are usually associated with the presence of a BEC. It is interesting that the physical realization of a BEC in a dilute gas has given impetus to the study of BEC for its own sake with a view of understanding many-body systems \cite{EJM96}. In addition, the confinement of photons and molecules in thermal equilibrium in an optical cavity reveals a BEC even for photons \cite{KS10}. The existence of superflow \cite{EK04} in solid helium $^4 \textup{He}$ has stimulated the search of a BEC in solid helium thus establishing the existence of BECs in all three states of matter--gas, liquid, and solid. The emergence of a self-organized supersolid phase, both a superfluid with crystalline order simultaneously, formed by a BEC coupled to an optical cavity has been observed \cite{BG10}. The quantum phase transition describing the supersolid, which is associated with a spontaneous broken spatial symmetry, is in quantitative agreement with the Dicke model of superradiance and is driven by an infinitely long-range interaction between the condensed atoms \cite{BG10}. A nonlocal potential is found to favor a crystalline BEC for the ground state of two-dimensional interacting bosons \cite{LLL11}. Numerical techniques have been used to predict a novel supersolid phase for an ensemble of Rydberg atoms in the dipole-blocked regime confined to two dimensions, interacting via a repulsive dipole potential softened at short distances \cite{CJB10}.  It is claimed that the superfluid droplet-crystal phase does not crucially depend on the dipolar form of the interaction at long distances \cite{CJB10}.

In this work, we use the Bogoliubov inequality to establish the properties of the interparticle potentials that allow for a BEC in crystalline order that can be regarded as a supersolid. It is shown that a local, two-particle potentials cannot result in a BEC with crystalline order in 2D and what is needed are nonlocal, two-body potentials of infinite range.

\section{Interacting Bose gas} Consider the Hamiltonian for an interacting Bose gas
\[
\hat{H}= \int d\textbf{r}\hat{\psi}^{\dag}(\textbf{r}) (\frac{-\hbar^2 }{2m} \nabla^2) \hat{\psi}(\textbf{r}) +  \int d\textbf{r}\hat{\psi}^{\dag}(\textbf{r}) V_{ext}(\textbf{r}) \hat{\psi}(\textbf{r})
\]
\begin{equation}
+ \int d\textbf{r}_{1}\int d\textbf{r}_{2} \int d\textbf{r}_{3} \int d\textbf{r}_{4} \hat{\psi}^{\dag}(\textbf{r}_{1})\hat{\psi}^{\dag}(\textbf{r}_{2}) V(\textbf{r}_{1}, \textbf{r}_{2},\textbf{r}_{3},\textbf{r}_{4}) \hat{\psi}(\textbf{r}_{4})\hat{\psi}(\textbf{r}_{3}),
\end{equation}
where $V_{ext}(\textbf{r})$ is a confining, external potential, $V(\textbf{r}_{1}, \textbf{r}_{2},\textbf{r}_{3},\textbf{r}_{4})$ is the two-particle interaction potential, and $\hat{\psi}(\textbf{r})$ and $\hat{\psi}^{\dag}(\textbf{r})$ are bosonic field operators that destroy or create a particle at spatial position $\textbf{r}$, respectively. The two-particle interaction potential $V(\textbf{r}_{1}, \textbf{r}_{2},\textbf{r}_{3},\textbf{r}_{4})$ must satisfy the following general conditions: (i) translational invariance, (ii) Galilean invariance, (iii) identical particles, (iv) rotational invariance, (v) space-reflection invariance, (vi) time-reversal invariance, and (vii) hermiticity \cite{MA71}. Therefore, in general,
\begin{equation}
V(\textbf{r}_{1}, \textbf{r}_{2},\textbf{r}_{3},\textbf{r}_{4}) = \delta(\textbf{r}_{1} + \textbf{r}_{2} -\textbf{r}_{3} -\textbf{r}_{4}) \langle \textbf{r}_{1} - \textbf{r}_{2}|V| \textbf{r}_{3} - \textbf{r}_{4}\rangle,
\end{equation}
which is referred to as a nonlocal potential. A mathematically simpler potential can be deduced from Eq. (2) if, in addition,
\begin{equation}
\langle \textbf{r}_{1} - \textbf{r}_{2}|V| \textbf{r}_{3} - \textbf{r}_{4}\rangle = \delta(\textbf{r}_{1} - \textbf{r}_{2} -\textbf{r}_{3} +\textbf{r}_{4}) V(|\textbf{r}_{1} - \textbf{r}_{2}|),
\end{equation}
and so
\begin{equation}
V(\textbf{r}_{1}, \textbf{r}_{2},\textbf{r}_{3},\textbf{r}_{4}) = \frac{1}{2} \delta(\textbf{r}_{1}  -\textbf{r}_{3}) \delta(\textbf{r}_{2}  -\textbf{r}_{4}) V(|\textbf{r}_{1} - \textbf{r}_{2}|),
\end{equation}
which is referred to as a local potential.

Macroscopic occupation in the single-particle state  $\psi(\textbf{r})$ result in the non-vanishing \cite{NNB60} of the quasi-average $\psi(\textbf{r}) = <\hat{\psi}(\textbf{r})>$ and so the boson field operator
\begin{equation}
\hat{\psi}(\textbf{r}) =\psi(\textbf{r}) + \hat{\varphi}(\textbf{r}),
\end{equation}
with
\begin{equation}
\psi(\textbf{r}) = \sqrt{\frac{N_{0}}{V(D)}}\sum_{\textbf{k}^\prime} \xi_{\textbf{k}^\prime} e^{i \textbf{k}^\prime \cdot\textbf{r}} \equiv\sqrt{\frac{N_{0}}{V(D)}} f(\textbf{r}) ,
\end{equation}
and
\begin{equation}
\sum_{\textbf{k}^\prime}|\xi_{\textbf{k}^\prime}|^2 =1,
\end{equation}
where $N_{0}$ is the number of atoms in the condensate and $V(D)$ is the D-dimensional ``volume" and $<\hat{\varphi}(\textbf{r})> =0$.  The operator $\hat{\varphi}(\textbf{r})$ has no Fourier components with momenta $\{\textbf{k}^\prime\}$ that are macroscopically occupied and so $\int d\textbf{r} \hat{\varphi}^\dag(\textbf{r}) \psi(\textbf{r}) = 0$. The separation of $\hat{\psi}(\textbf{r})$ into two parts gives rise to the following (gauge invariance) symmetry breaking term in the Hamiltonian (1)
\[
\hat{H}_{symm} =  \int d\textbf{r}_{1} \hat{\varphi}^\dag(\textbf{r}_{1}) \int d\textbf{r}_{2}\int d\textbf{r}_{3}\int d\textbf{r}_{4} \psi^*(\textbf{r}_{2}) [ V(\textbf{r}_{1}, \textbf{r}_{2},\textbf{r}_{3},\textbf{r}_{4}) + V(\textbf{r}_{2}, \textbf{r}_{1},\textbf{r}_{3},\textbf{r}_{4}) ]\psi(\textbf{r}_{3})\psi(\textbf{r}_{4})   + h. c.
 \]
\begin{equation}
\equiv \int d\textbf{r}_{1} \hat{\varphi}^\dag(\textbf{r}_{1}) \chi(\textbf{r}_{1}) + h. c.
\end{equation}

The presence of this nonzero $\hat{H}_{symm}$ in the Hamiltonian gives rise to further macroscopic occupation in states other than the original state given by $\psi(\textbf{r})$ and so the condensate wavefunction $\psi(\textbf{r})$ gets modified by augmenting the single-particles states where macroscopic occupation occurs. In such a case, macroscopic occupation in the state $b$ would give rise to macroscopic occupation in the states $a$, such that $a\neq b$, whenever  the matrix element $<ab|\hat{V}|bb>$  of the potential $\hat{V}$, which is the last term in Eq. (1), does not vanish.  For instance, macroscopic occupation only in the single-particle state with momentum $\textbf{p}$, which corresponds to a uniform condensate, does not give rise to macroscopic occupation in any other momentum state since the matrix element in the momentum representation $<\textbf{q}\textbf{p}| \hat{V}|\textbf{p}\textbf{p}>$ vanishes by momentum conservation unless $\textbf{q}= \textbf{p}$. This consistency proviso requires that the correct condensate wavefunction $\psi(\textbf{r})$ corresponds to that which gives rise to no symmetry breaking term in the Hamiltonian. That is to say, $\hat{H}_{symm}$ vanishes for the correct condensate wavefunction $\psi(\textbf{r})$.

For instance, macroscopic occupation in the single-particle states with momenta $\textbf{k}, \textbf{k} \pm\textbf{q}_{1}, \textbf{k} \pm\textbf{q}_{2} $ gives rise, with the aid of the symmetry breaking term $\hat{H}_{symm}$ and owing to linear momentum conservation, to additional macroscopic occupation in single-particle momenta states. Therefore, for $\hat{\varphi}^\dag(\textbf{r})$ to be orthogonal to both $\psi(\textbf{r})$ and $\chi(\textbf{r})$ and so $\hat{H}_{symm}=0$, one must have macroscopic occupation in all the momentum states $\textbf{k} + n_{1}\textbf{q}_{1} + n_{2}\textbf{q}_{2}$, with $n_{1}, n_{2} = 0, \pm 1, \pm 2, \cdots.$   Accordingly,
\begin{equation}
\psi_{\textbf{k}}(\textbf{r}) = \sqrt{\frac{N_{0}}{V(D)}} \sum_{n_{1},  n_{2} = - \infty}^{\infty}  \xi_{\textbf{k}+ n_{1} \textbf{q}_{1} + n_{2} \textbf{q}_{2} } \hspace{0.1in} e^{i (\textbf{k}+ n_{1} \textbf{q}_{1} + n_{2} \textbf{q}_{2} )\cdot\textbf{r}} \equiv  e^{i \textbf{k}\cdot \textbf{r}} u_{\textbf{k}}(\textbf{r}).
\end{equation}
Note that the BEC (9) generates a real-space, crystalline distribution of atoms since $u_{\textbf{k}}(\textbf{r}) = u_{\textbf{k}}(\textbf{r} +\textbf{t}_{m})$ for any primitive lattice translation vector $\textbf{t}_{m}$ with $e^{\textbf{q}_{i}\cdot \textbf{t}_{m}} =1$ for $i=1, 2$ and so (9) is of the Bloch form. For D = 2, the vector $\textbf{t}_{m} = m_{1} \textbf{a} + m_{2} \textbf{b}$, where $m_{1}$ and $m_{2}$ can take all integer values and $\textbf{a}$ and $\textbf{b}$ are the edges of the unit cell, which form parallelograms given by the five Bravais lattices. Note that $\textbf{a} \cdot \textbf{q}_{1} =2 \pi$,  $\textbf{a} \cdot \textbf{q}_{2}= 0$,  $\textbf{b} \cdot \textbf{q}_{1} = 0$, and  $\textbf{b} \cdot \textbf{q}_{2}=2 \pi$.

\section{ Bogoliubov inequality} The absence or presence of a BEC in spatial dimensions $D \leq 2$ is based on Bogoliubov's inequality
\begin{equation}
\frac{1}{2}\langle\{\hat{A},\hat{A}^{\dag}\}\rangle \geq k_{B}T |\langle[\hat{C},\hat{A]}\rangle|^2 / \langle[[\hat{C}, \hat{H}], \hat{C}^{\dag}]\rangle,
\end{equation}
where $\hat{H}$ is the Hamiltonian (1) of the system, the brackets denote thermal averages, and the operators $\hat{A}$ and $\hat{C}$ are arbitrary provided all averages exist. Consider first the case that $\hat{H}$ has arbitrary local interparticle and external potentials, that is, $V(\textbf{r}_{1}, \textbf{r}_{2},\textbf{r}_{3},\textbf{r}_{4})$ is given by Eq. (4).

Consider the following operators,
\begin{equation}
\hat{C} = \int d\textbf{r} e^{i\textbf{k}\cdot \textbf{r}} \hat{\psi}^{\dag}(\textbf{r})\hat{\psi}(\textbf{r})
\end{equation}
and
\begin{equation}
\hat{A} = \int d \textbf{r} \int d \textbf{r}^{\prime} e^{-i\textbf{k}\cdot \textbf{r}} f(\textbf{r}) f^{\ast}(\textbf{r}^\prime) \hat{\psi}^{\dag}(\textbf{r})\hat{\psi}(\textbf{r}^\prime),
\end{equation}
where $\textbf{k}$ is arbitrary. Now,
\begin{equation}
\langle [A^\dag,A]\rangle = V(D)\langle[\hat{C},\hat{A}]\rangle = V^2(D)\{ N_{0} - N_{0} |A_{\textbf{k}}|^2 -\sum_{\textbf{q}}\langle \hat{a}^\dag_{\textbf{q}} \hat{a}_{\textbf{q}}\rangle  | \xi_{\textbf{q}+ \textbf{k}}|^2 \},
\end{equation}
\begin{equation}
\langle[[\hat{C}, \hat{H}], \hat{C}^{\dag}]\rangle = \frac{\hbar^2 k^2}{m} N,
\end{equation}
with
\begin{equation}
A_{\textbf{k}} = \sum_{\textbf{k}^\prime} \xi_{\textbf{k}^\prime} \xi^{*}_{\textbf{k}^\prime + \textbf{k}}
= \frac{1}{N_{0}} \int \textup{d} \textbf{r} |\psi(\textbf{r})|^2 e^{i \textbf{k}\cdot \textbf{r}}
\end{equation}
with the aid of Eq. (6), where $N$ is the total number of particles. Now, $|A_{\textbf{k}}| \leq 1$  by the Cauchy-Schwarz inequality  where the equality holds for $\textbf{k} = \textbf{0}$, that is, $A_{0}=1$ with the aid of Eq. (7). The vector $\textbf{q} \notin \{\textbf{k}^\prime\}$, where $\{\textbf{k}^\prime\}$ is the set of condensate vectors for which $\xi_{\textbf{k}^\prime} \neq 0$.  For a BEC at rest, $|\xi_{\textbf{k}}|^2 = |\xi_{- \textbf{k}}|^2$. Note that $\xi_{\textbf{q} + \textbf{k}}\neq 0$ for $(\textbf{q} + \textbf{k}) \in \{\textbf{k}^\prime\}$ and $\xi_{\textbf{q} + \textbf{k}}= 0$ for $\textbf{q}\notin \{\textbf{k}^\prime\}$ and $ \textbf{k} \in \{\textbf{k}^\prime\}$.

It is important to remark that Eq. (14) holds for local interparticle potentials but need not hold true for the general case of nonlocal potentials \cite{MA71}. Therefore, for a BEC to exist in spatial dimensions $\textup{D} \leq 2$, the $1/k^2$-singularity that results from the Bogoliubov commutator in the denominator of Eq. (10) must me removed. For local interparticle potentials this can be accomplished only provided the origin $\textbf{k} =\textbf{0}$ is a limit point (or point of accumulation) of condensates. This would correspond to both momenta $\textbf{q}_{1}$ and $\textbf{q}_{2}$ in Eq. (9) approaching zero. Of course, in this case the BEC (9) ceases to be of the Bloch form, albeit, remaining nonuniform. The case where the interparticle potential is nonlocal is discussed in Section IV.

We sum the Bogoliubov inequality (10) over the single-particle momentum states in the set $\{\textbf{k}^\prime\}$ constituting the condensate, which includes an arbitrary neighborhood of the point of accumulation of the condensate at $\textbf{k}^\prime = \textbf{0}$ that corresponds to a condensate at rest. We want to find an upper bound of the anticommutator   $\langle\{\hat{A},\hat{A}^{\dag}\}\rangle = 2\langle \hat{A}\hat{A}^{\dag}\rangle + \langle [\hat{A}^{\dag},\hat{A}]\rangle$. We extend the sum over the first term $\langle \hat{A}\hat{A}^{\dag}\rangle$ over all values of $\textbf{k}$ thus obtaining a larger upper bound
\begin{equation}
M N_{0}N V^2(D) \geq \sum_{\textbf{k}}\langle \hat{A}\hat{A}^{\dag}\rangle,
\end{equation}
where the condensate wavefunction $\psi(\textbf{r})$ is orthogonal to the operator $\hat{\varphi}^\dag(\textbf{r})$ and we assume that the condensate density is bounded from above by $|f(\textbf{r})|^2 \leq M$ with $1 \leq M <\infty$ since $\int d \textbf{r}|f(\textbf{r})|^2 =V(D)$. In (16) use has been made of the completeness relation for the momentum eigenstates and a negative term resulting from a single commutation has been dropped. Note that we are considering a condensate where all the single-particle states with momentum $\textbf{k}^\prime$ are occupied macroscopically with $\textbf{k}^\prime = \textbf{0}$ a point of accumulation. In addition, we are supposing that the number of particles in the ``volume" $V(D)$ is fixed, that is, we are employing a canonical ensemble and so  $\int d\textbf{r}\hat{\psi}^{\dag}(\textbf{r})\hat{\psi}(\textbf{r}) = \sum_{\textbf{k}} \hat{a}^\dag_{\textbf{k}} \hat{a}_{\textbf{k}} =  \hat{N}$ is actually the c-number $N$.

Consider next the sum over $\textbf{k}^\prime$ of the commutator $\langle[\hat{A}^\dag,\hat{A]}\rangle$,
\begin{equation}
\sum _{\textbf{k}^\prime} \langle[\hat{A}^\dag,\hat{A]}\rangle = N_{0} V^2(D)  \sum _{\textbf{k}^\prime} [ 1 - |A_{\textbf{k}^\prime}|^2],
\end{equation}
with the aid of (13) and where $\xi_{\textbf{q} + \textbf{k}} =0$ for $\textbf{q}\notin \{\textbf{k}^\prime\}$ and $\textbf{k}\in \{\textbf{k}^\prime\}$. This sum over the commutator is bounded from above provided the sum is restricted to values of $\textbf{k}^\prime $ that have a finite, upper bound. Now the right-hand side (RHS) of inequality (10) becomes
\begin{equation}
k_{B}T |\langle[\hat{C},\hat{A]}\rangle|^2 / \langle[[\hat{C}, \hat{H}], \hat{C}^{\dag}]\rangle = \frac{m k_{B}T}{\hbar^2} \frac{ V^2(D) N_{0}^2}{N} \sum_{\textbf{k}^\prime}\Bigl ( \frac{1 - |A_{\textbf{k}^\prime}|^2}{k^\prime} \Bigr )^2
\end{equation}
with the aid of Eqs. (13) and (14). Note that the RHS is bounded in the upper limit of the sum; however, it is the lower limit as $\textbf{k}^\prime \rightarrow \textbf{0}$ for $D \leq 2$ where the sum may diverge which would result in no BECs, viz., $N_{0}=0$ for $T>0$. Combining Eqs. (16)--(18), we have for the Bogoliubov inequality,
\begin{equation}
M N + \frac{1}{2} \sum_{\textbf{k}^\prime}( 1 - |A_{\textbf{k}^\prime}|^2)  \geq \frac{m k_{B}T }{\hbar^2} \frac{ N_{0}}{N} \sum_{\textbf{k}^\prime}\Bigl( \frac{1 - |A_{\textbf{k}^\prime}|^2}{k^\prime} \Bigr) ^2.
\end{equation}
 Therefore, the existence of a BEC for $T>0$ requires the convergence of the sum on the RHS of (19) over the macroscopically occupied single-particle momentum states $\textbf{k}^\prime$ of the condensate. Note that the sums in (19) over the condensate momenta can be approximated by integrals according to $\sum_{\textbf{k}^\prime} \rightarrow V(D) \int \textup{d} \textbf{k}^\prime$ and so
 \begin{equation}
M \frac{ N}{V(D)} + \frac{1}{2} \int \textup{d} \textbf{k}^{\prime}( 1 - |A_{\textbf{k}^\prime}|^2)  \geq \frac{m k_{B}T }{\hbar^2} \frac{ N_{0}}{N} \int \textup{d} \textbf{k}^{\prime} \Bigl( \frac{1 - |A_{\textbf{k}^\prime}|^2}{k^\prime} \Bigr) ^2.
 \end{equation}
 The integral on the RHS has no infrared divergence since by (15), $(1 -|A_{\textbf{k}^\prime}|^2)$ vanishes quadratically as $\textbf{k}^\prime \rightarrow  \textbf{0}$ and so the $1/k^2$--singularity is removed thus allowing the existence of a BEC for $D\leq 2$.

\section{Supersolid BEC} The analysis of the possible existence of BEC for $D\leq 2 $ in the previous section was based on two-particle, local interactions. Clearly, local interparticle potentials cannot give rise to a BEC of the Bloch form for $D=2$ since the $1/k^2$--singularity in the Bogoliubov inequality (10), owing to the $k^2$--behavior of the Bogoliubov  commutator (14), cannot be removed and still preserve the Bloch form for the BEC. For local interparticle potentials, the Bogoliubov commutator is given solely by the kinetic energy term of the Hamiltonian (1). This need not be so for nonlocal, long-range interparticle potentials \cite{MA71}.

The $k^2$--behavior of the double commutator (14) follows from the kinetic energy term of the Hamiltonian since local interparticle potentials $\hat{V}$ do not contribute to the Bogoliubov commutator, viz., $\langle[[\hat{C}, \hat{V}], \hat{C}^{\dag}]\rangle = 0$. It is interesting that the latter is not the case for nonlocal potentials \cite{MA71}. If, for instance, the two-particle potential is a sum of a local and a nonlocal potential, then the former potential does not contribute to the Bogoliubov commutator while the latter does and if the decay of the nonlocal potential with distance is sufficiently slow, then $\langle[[\hat{C}, \hat{V}], \hat{C}^{\dag}]\rangle \propto k^{2-\epsilon} $ with $\epsilon > 0$ as $k\rightarrow 0$ \cite{MA71}. Accordingly, the behavior of the Bogoliubov commutator in the limit $k\rightarrow 0$ is dominated by the nonlocal term rather than the kinetic energy term as is the case in the local case. Therefore, the symmetry breaking term (8) allows a BEC of the Bloch form for $D=2$ provided the potential between the condensate atoms are given by an infinitely long-range nonlocal potential.

Recently, a Dicke quantum phase transition was realized in an open system formed by a BEC coupled to an optical cavity that gives rise to a self-organized supersolid phase \cite{BG10}. It is interesting that the phase transition is driven by infinitely long-range interactions between the condensed atoms. The analogy of that work to the Dicke model is based on the interaction Hamiltonian that gives rise to a coupling of the pump and cavity fields to the zero-momentum states of the atoms to the symmetric superposition of atomic states that carry an additional unit of photon momentum. This is quite analogous to our dynamically generated symmetry breaking term that allows condensation in atomic states that are integer multiples of a given condensate momentum. Note, however, that only one additional unit of photon momentum along the $x$ and $z$ directions are consider \cite{BG10}. Higher-order momentum states must be included in order to describe atomic localization at the sites of the emergent checkerboard pattern, that is, to represent a true BEC Bloch form \cite{BG10}.

\section{Summary and conclusion}

We use the Bogoliubov inequality to establish necessary conditions for the existence of two-dimensional BECs of the Bloch form and show that the interparticle potential must be of sufficiently long range and nonlocal. It has been shown experimentally that such a quantum phase transition occurs in an almost pure $^{87}\textup{Rb}$ gas resulting in the onset of self-organization into a supersolid \cite{BG10}. This behavior is equivalent to a dynamical version of the normal-to-superradiant quantum phase transition of the Dicke model \cite{BG10}.

\begin{newpage}
\bibliography{basename of .bib file}

\begin{thebibliography}{8}
\bibitem{EJM96} J. R. Ensher, D. S. Jin, M. R. Matthews, C. E. Wieman, and E. A. Cornell, Phys. Rev. Lett. \textbf{77}, 4984 (1996).
\bibitem{KS10} J. Klaers, J. Schmitt, F. Vewinger, and M. Weitz, Nature (London) \textbf{468}, 545 (2010).
\bibitem{EK04} E. Kim and M. H. W. Chan, Nature (London) \textbf{427}, 225 (2004).
\bibitem{BG10} K. Baumann, C. Guerlin, F. Brennecke, and T. Esslinger  K. Baumann \emph{et al.}, Nature (London) \textbf{464}, 1301 (2010).
\bibitem{LLL11} X. Li, W. V. Liu, and C. Lin, Phys. Rev. A \textbf{83}, 021602(R) (2011).
\bibitem{CJB10} F. Cinti, P. Jain, M. Boninsegni, A. Micheli, P. Zoller, and G. Pupillo, Phys. Rev. Lett. \textbf{105}, 135301 (2010).
\bibitem{MA71} M. Alexanian, Phys. Rev. A \textbf{4}, 1684 (1971).
\bibitem{NNB60} N. N. Bogoliubov, Physica (Amsterdam) \textbf{26}, S1 (1960).




\end{thebibliography}

\end{newpage}

\end{document}